\documentclass[12pt,a4paper]{article}
\usepackage{amsthm}
\usepackage{amsmath}
\usepackage{natbib}
\usepackage[colorlinks,citecolor=blue,urlcolor=blue,filecolor=blue,backref=page]{hyperref}
\usepackage{graphicx}
\usepackage{xcolor}
\usepackage{algorithm}
\usepackage{adjustbox}
\usepackage[noend]{algpseudocode}
\usepackage{natbib}
\usepackage{bbm, dsfont}
\usepackage{amsfonts}
\usepackage{siunitx}
\usepackage[paperheight=13in,paperwidth=9.5in]{geometry}

\newcommand{\be}{\begin{equation}}
\newcommand{\ee}{\end{equation}}
\newtheorem{remark}{\textit{Remark}}

\begin{document}

\title{\textbf{\Large On the Estimation of Population
		Size from a Dependent Triple Record System}}

\date{}
\maketitle
\author{
\begin{center}
\vskip -1cm

Kiranmoy Chatterjee \\
Interdisciplinary Statistical Research Unit, Indian Statistical Institute\\
kiranmoy07@gmail.com\\

Prajamitra Bhuyan\\
Department of Mathematics, Imperial College London\\
bhuyan.prajamitra@gmail.com
\end{center}
}

\begin{abstract}
Population size estimation based on capture-recapture experiment under triple record system is an interesting problem in various fields including epidemiology, population studies, etc. In many real life scenarios, there exists inherent dependency between capture and recapture attempts. We propose a novel model that successfully incorporates the possible dependency and the associated parameters possess nice interpretations. We provide estimation methodology for the population size and the associated model parameters based on maximum likelihood method. The proposed model is applied to analyze real data sets from public health and census coverage evaluation study. The performance of the proposed estimate is evaluated through extensive simulation study and the results are compared with the existing competitors. The results exhibit superiority of the proposed model over the existing competitors both in real data analysis and simulation study.
\end{abstract}


\section[]{Introduction}
\label{sec:intro}
Estimation of population size or the number of vital events occurred, during a given time span, is a relevant statistical problem in various scientific disciplines including epidemiology, population studies, and life sciences. Federal agencies are generally interested in such estimates for planning and policy formulation. In general, census or any registration system often fails to capture all the individuals and that leads to undercoverage of the population under consideration. However, in some instances, duplicate records or members outside the target population are included in the census or any other registers because of erroneous enumeration. This issue is known as overcoverage, and it is a common practice to identify and remove the erroneous inclusions through administrative follow-up actions \citep{Chipperfield17} or adjust the census data based on the estimate of overcoverage rate \citep{Zhang15}. In this paper, we only focus on the issues related to the commonly encountered problem of undercoverage assuming that the available data are free from any erroneous inclusion. In order to reduce the undercoverage error, information from more than one attempt needs to be considered. The data obtained from various sources are summarized by matching the lists of captured individuals and analyzed to obtain an estimate of the unknown population size \citep{Rastogi12}. Specifically for the human population, this data collection technique from multiple capture attempts is popularly known as Multiple Record System (MRS), which is similar to the capture-mark-recapture (CMR) technique, widely used in Ecology \citep{Otis78, Seber86}. After the pioneering work by \citet{Chandrasekar49}, CMR type technique became popular for assessing the undercoverage of census or demographic registrar \citep{Ayhan00, Wachter08} and under-ascertainment in traditional epidemiological surveillance \citep{Chao01a, Dreyfus14}. Several models with suitable assumptions on capture probabilities have been proposed for estimation of the population size under MRS \citep{Otis78}. In practice, more than three sources are seldom used for human population due to various practical constraints.

The most common model in applications involving human populations assumes that capture probabilities across individuals are same in each attempt (i.e. \textit{homogeneity}) but different over the various sources (i.e. \textit{time variation}). This model is known as  $M_t$ model, which further assumes that capture statuses are causally independent over different capture attempts \citep{Otis78, Wolter86}. However, in many situations, the capture status of an individual in a list may be dependent on his/her capture statuses in the previous attempts. This phenomena is known as \textit{list-dependence} (or \textit{local dependence}) in the literature \citep{Chao01a}. On the other hand, capture probability in each list may vary across individuals, which is typically known as \textit{heterogeneity (or individual heterogeneity)} in capture probabilities \citep{Wolter86}. \citet{Chao01a} discussed the fact that even if the individual capture statuses are independent across different capture attempts, the lists may become dependent whenever the capture probabilities are heterogeneous across individuals in both the lists. These two types of dependencies are usually confounded and not identifiable individually. Any of these two dependencies leads to a bias in the resulting estimate based on the $M_t$ model, and this is popularly known as \textit{correlation bias} in the domain of census undercount estimation \citep{Fay88, Chatterjee16b}. However, it is noted that the estimate would not be biased if such heterogeneity exists only in one of the two causally independent lists \citep{Chao01a,Heijden18}. To reduce the extent of heterogeneity as well as the dependence induced by heterogeneity in the population, it is often desirable to employ post-stratification of the population based on appropriate demographic (e.g., age, race, sex) and geographic variables \citep{Wolter86,Islam15}. However, the list-dependence may still exist in the resulting post-stratified data. Ample amount of research work on modeling of list-dependence under MRS are available in the literature \citep{Chao01a, Fienberg72a}. \citet{Fienberg72a} discussed log-linear models which incorporate possible dependencies among the capture attempts for multiple recapture census. Interested readers are referred to \citet{Fienberg72a, Fienberg72b,Bishop75,WorkingGroup95} for details. However, it is important to note that the parameters associated with the log-linear models are not well-interpretable to the practitioners \citep{Coumans17}. \citet{Otis78} proposed an alternative model, popularly known as $M_{tb}$, which accounts for the list-dependence with respect to behavioral response variation. In this context, \citet{Chao00} discussed likelihood based methods for estimation of the population size under some assumptions on capture probabilities for more than two capture attempts. This model possesses several practical limitations (\textit{see} Subsection \ref{ecological} for more details).  

As discussed before, two or three capture attempts are commonly used for the human population. Dual system or Dual-record System (DRS) is a particularization of MRS, where number of capture attempts is only two \citep{Chao01a,Chatterjee16c}. Interestingly, the lists from two different sources in DRS are often correlated. For examples, (\textit{i}) laboratory and hospital records for infectious diseases \citep{Hest02} and (\textit{ii}) records from census and post-enumeration survey for assessing coverage error in the census \citep{Zaslavsky93} are often dependent. Therefore, the estimate based on $M_t$ model would be biased. Moreover, DRS is not sufficient to infer about this interdependence between the two sources \citep{Chao01a}. In such a situation, one can post-stratify the population and model the possible interdependence under suitable assumptions on the dependence structures of the strata \citep{Wolter90, Chatterjee17}. Alternatively, an additional source of data can be added to cover up more eligible events and assess the underlying interdependence among the lists. Such data structure from three separate capture attempts is known as Triple-record System (TRS). This data structure is commonly used in epidemiology and demography for a wide range of studies including enumeration of hard-to-count population or rare events \citep{Tsay01, Ruche13, Dreyfus14}. Census undercount study is another field where TRS based estimates are less affected by correlation bias \citep{Griffin14}. Triple system estimation involving an administrative record as a third source, along with the two sources - census and post-enumeration survey, can be used to assess various pair-wise dependencies among the lists.

In this article, we propose a novel Trivariate Bernoulli Model (TBM), which incorporates the inherent interdependence among capture attempts in TRS. In addition, this model possesses easy interpretations compared to the existing competitors. Capture attempts (or lists) in the census undercoverage study are commonly \textit{time-ordered} \citep{Zaslavsky93}. These attempts are organized within a short span of time so that the basic assumption of \textit{the closed population} (See Section \ref{TRS}) holds. As discussed before, individual heterogeneity in capture probability is contained by suitable post-stratification of the target population. However, in time-ordered scenarios, capture probabilities may vary across the individuals depending on the capture statuses in the previous attempts. This type of heterogeneity associated with behavioral response variation results only due to the list-dependence \citep{Chao01a}. In this context, several ecological models has been proposed in the literature under the assumption of `time-variation' \citep{Chao00,Griffin14}. On the other hand, some studies (e.g. epidemiological surveillance) commonly involve capture attempts, which are not according to an ordered sequence with time or the order may vary across individuals. In such cases, any model involving behavioral response variation or any estimator depending on the time-ordered list has limited use in practice \citep{Chao01a}. Therefore, we propose two sub-models from TBM in this paper for accounting these two aforementioned scenarios. Moreover, based on extensive simulation study, we demonstrate that our proposed models and associated estimates exhibit superiority in terms of relative root mean square (RRMSE) over the existing models in the literature. We first describe the Triple Record System and related existing models and their limitations in Section \ref{TRS} and \ref{existing}, respectively. In Section \ref{proposed}, we propose TBM and discuss associated estimation methodology. The proposed method is illustrated with the analysis of real datasets in Section \ref{real_data_analysis}. In Section \ref{Simulation}, we compare the performance of the proposed estimators with the existing competitors. Finally, we end with some concluding remarks in Section \ref{conclusion}.


\section{Triple Record System}\label{TRS}
Let us consider a population \emph{U} of unknown size \emph{N} where three attempts are organized to enumerate all individuals in the population. Now, we consider two basic assumptions: (\textit{I}) population is closed during the time of enlisting people by the three capture attempts, (\textit{II}) individuals are homogeneous with respect to their capture probabilities in each of the three attempts. Therefore, by matching the individuals enlisted in the three different surveys, the available data is presented in a incomplete $2^3$ table (\textit{see} Table \ref{Tab:1}). Let us denote the capture status of an individual in the first, second and third lists by $i,j$ and $k$, respectively. The dummy variables $i,j$ and $k$ take value 1 for a capture and 0 otherwise. The total number of individuals having a particular capture status, say ($i,j,k$), is denoted by $x_{ijk}$, and the associated cell probability is denoted by $p_{ijk}$ (\textit{see} Table \ref{Tab:1}). Note that the unknown number of individuals, not captured by any source, denoted by $x_{000}$, makes the population size \emph{N}($=x_{\cdot\cdot\cdot}$) unknown.  
\begin{table}[h]
	\begin{center}
		\caption{Data structure corresponds to Triple Record System with associated cell probabilities in [ ] and $x_{\cdot\cdot1}+x_{\cdot\cdot0}=N$}
		\begin{tabular}{lcccccccc}
			\hline
			&\multicolumn{8}{c}{List 3} \\
			\cline{2-9}
			&\multicolumn{3}{c}{In}&&&\multicolumn{3}{c}{Out} \\			
			\cline{2-9}
			&\multicolumn{3}{c}{List 2}&&&\multicolumn{3}{c}{List 2} \\	
			List 1 & In & Out & Total &&& In & Out & Total\\
			\hline \hline
			In & $x_{111} [p_{111}]$ & $x_{101} [p_{101}]$ & $x_{1\cdot1} [p_{1\cdot1}]$ &&& $x_{110} [p_{110}]$ & $x_{100}[p_{100}]$ & $x_{1\cdot0} [p_{1\cdot0}]$\\
			Out& $x_{011}[p_{011}]$ & $x_{001}[p_{001}]$ & $x_{0\cdot1} [p_{0\cdot1}]$ &&& $x_{010}[p_{010}]$ & \textbf{$x_{000} [p_{000}]$} & $x_{0\cdot0} [p_{0\cdot0}]$\\ 
			\hline
			Total& $x_{\cdot11} [p_{\cdot11}]$ & $x_{\cdot01} [p_{\cdot01}]$ & $x_{\cdot\cdot1} [p_{\cdot\cdot1}]$ &&& $x_{\cdot10} [p_{\cdot10}]$ & $x_{\cdot00} [p_{\cdot00}]$ & $x_{\cdot\cdot0} [p_{\cdot\cdot0}]$\\
			\hline
		\end{tabular}
		\label{Tab:1}
	\end{center}
\end{table}

\section{Existing Models \& Estimates}\label{existing}
As discussed before, $M_t$ model assumes the causal independence among the lists for estimation of $N$. However, the causal independence assumption among the available lists is criticized in the context of public health and census coverage study (\textit{see} \citeauthor{Chao01a}, \citeyear{Chao01a}; \citeauthor{Zaslavsky93}, \citeyear{Zaslavsky93}). The major concern is that an individual's capture status in the first attempt may affect its capture status in the subsequent attempts leading to positive or negative dependence among the lists. In the following subsections, we will briefly discuss $M_{tb}$ model and some relevant log-linear models in this context. 

\subsection{Time-Behavioral Response Variation Model: $M_{tb}$}\label{ecological}
It is often observed in MRS that an individual's behavior changes with the time of subsequent recapture attempts after the initial attempt. An individual who is captured by the first attempt may have more (or less) chance to be included in the second attempt than the individual who has not been captured by the first attempt. This change is known as \textit{behavioral response variation} \citep{Otis78, Chatterjee16c}. When this \textit{behavioral response variation} is considered along with the assumption of \textit{time variation} in capture probabilities, one would have a model known as $M_{tb}$ model \citep{Wolter86}. However, the model parameters are not estimable as well as identifiable in DRS under any circumstances. To obtain an estimate of $N$, three or more capture attempts under suitable assumptions are needed. Now we briefly discuss the parameterization and associated likelihood of $M_{tb}$ model in TRS.

Let us denote the first-time capture probability of any individual in the $l$-th list by $f_l$ for $l=1, 2, 3$, whereas the recapture probability is denoted by $c_l$ for $l=2, 3$. Further, $u_l$ and $m_l$ denote, respectively, the number of first-time captured and recaptured individuals in the $l$-th list. In this case, the minimal sufficient statistic ($u_1=x_{1\cdot\cdot},u_2=x_{01\cdot},u_3=x_{001},m_2=x_{11\cdot}, m_3=x_{101}+x_{011}+x_{111}$) fails to estimate the $M_{tb}$ model comprising six parameters including $N$. In order to overcome this difficulty, recapture probabilities and first-capture probabilities are assumed to be related by a \textit{constant proportional parameter}, i.e., $c_l/f_l=\phi,$ for $l=2,3$. For details about the estimability issues associated with this model, readers are referred to \citet{Chao00}. The model along with this assumption will be called $M_{tb}$ model hereafter. Letting $\textbf{\textit{f}}=(f_1,f_2,f_3)$, the likelihood function under the above assumption is given by
\begin{eqnarray}
L(N,\textbf{\textit{f}},\phi)=\frac{N!}{(N-x_0)}f_1^{u_1}(1-f_1)^{N-u_1}\phi^{m_2+m_3}\prod_{l=2}^{3}f_l^{u_l+m_l}(1-f_l)^{N-M_{l+1}}(1-\phi p_l)^{M_l-m_l},\nonumber
\end{eqnarray}
where $x_0=\sum_{i,j,k: ijk\neq000}x_{ijk}$ and $M_l=u_1+u_2+...+u_{l-1}$ denote the total number of distinct captured individuals and the number of individuals captured at least once prior to the $l$-th attempt, respectively. \citet{Chao00} discussed various likelihood based methods for estimation of $N$ and other associated model parameters. It is important to note that the $M_{tb}$ model is basically designed to analyze data structures commonly found in capture-recapture experiments for wildlife population. It is clearly seen that $M_{tb}$ model does not utilize full information available from seven known cells of the incomplete $2^3$ table associated with the TRS (\textit{see} Table \ref{Tab:1}). Hence, it looses efficiency in order to estimate $N$ under the existence of various dependence structures among three capture attempts (\textit{see} RRMSEs in Tables \ref{Tab:3} and \ref{Tab:4}). In addition to that, the assumption of `constant proportional parameter' (i.e., $c_2/f_2=c_3/f_3=\phi$) is not justified in the existing literature for human population.

\subsection{Log-linear Models}\label{log-linear}
\citet{Fienberg72b} discussed log-linear models for estimation of the population size $N$ incorporating dependence among the capture attempts using interaction effects. Here, we briefly describe the relevant log-linear models (LLMs) for TRS. Let $m_{ijk}$ be the expected number of individuals corresponding to the ($i,j,k$)th cell in Table \ref{Tab:1}. The general log-linear model under TRS is given by
\begin{eqnarray}
\log(m_{ijk})=u_0 + u_{1(i)} + u_{2(j)} + u_{3(k)} + u_{12(ij)} + u_{13(ik)} + u_{23(jk)} + u_{123(ijk)},\label{log-linear_gen}
\end{eqnarray} 
where $u_{l(0)}+u_{l(1)}=0$,  $u_{l l^{\prime} (0j)}+u_{l l^{\prime}(1j)}=0$,
$u_{l l^{\prime} (i0)}+u_{l l^{\prime}(i1)}=0$,
$u_{l l^{\prime}l^{\ast} (0jk)}+u_{l l^{\prime}l^{\ast}(1jk)}=0$,
$u_{l l^{\prime}l^{\ast} (i0k)}+u_{l l^{\prime}l^{\ast}(i1k)}=0$,
$u_{l l^{\prime}l^{\ast} (ij0)}+u_{l l^{\prime}l^{\ast}(ij1)}=0$,
for all $l, l^{\prime},l^{\ast}=1,2,3$, with $l\ne l^{\prime}\ne l^{\ast}$.
The parameters $u_l$, $u_{l l^{\prime}}$ and $u_{123}$ denote the main effects, pairwise interaction effects and the second order interaction effect for $l=1,2,3$, and $l^{\prime}=1,2,3$. In the log-linear model under capture-recapture setting, observation $x_{ijk}$ is assumed to be a realization of independent Poisson random variate with expectation $m_{ijk}$ for all $(i,j,k)$th cells, except the $(0,0,0)$th cell. See \citet{Fienberg72a, Fienberg72b} and \citet{WorkingGroup95} for more details.

Under the assumption $u_{123}=0$ (i.e., no second order interaction in (\ref{log-linear_gen})), we  refer the reduced log-linear model as LLM-1, and the associated MLE for $m_{000}$ is given by
\begin{eqnarray}
\hat{m}_{000}^{(1)}=\frac{x_{111}x_{001}x_{100}x_{010}}{x_{101}x_{011}x_{110}}. \nonumber
\label{est_no_2nd_order}
\end{eqnarray}
This is equivalent to the estimator proposed by \citet[p. 285]{Zaslavsky93} with $\alpha_{EPA}=1$.

In a series of time-ordered capture attempts (i.e. when the counting surveys are organized in an ordered sequence w.r.t time), often it becomes reasonable to assume that first order interaction effect between List-1 and List-2 ($u_{12}$) and the same between List-2 and List-3 ($u_{23}$) may have significant impact while the interaction between first and third lists ($u_{13}$) is null, in addition to the primary assumption of $u_{123}=0$. Under these assumptions, the model (\ref{log-linear_gen}) reduces to
\begin{eqnarray}
\log(m_{ijk})=u_0 + u_{1(i)} + u_{2(j)} + u_{3(k)} + u_{12(ij)} + u_{23(jk)},\label{log-linear_two_1st_order}
\end{eqnarray}
and we refer to the log-linear model in (\ref{log-linear_two_1st_order}) as LLM-2. The MLE of $m_{000}$ under this model is given by
\begin{eqnarray}
\hat{m}_{000}^{(2)}&=&\frac{x_{001}x_{100}}{x_{101}}. \nonumber
\label{est_two_1st_order}
\end{eqnarray}
Finally, estimates of $N$ are given by $\hat{N}_{LLM}^{(1)}=x_0+\hat{m}_{000}^{(1)}$ and $\hat{N}_{LLM}^{(2)}=x_0+\hat{m}_{000}^{(2)}$ under LLM-1 and LLM-2, respectively. For details on the derivation of estimates  and their respective asymptotic variance  under LLM-1 and LLM-2, readers are suggested to refer \citet{Fienberg72b}.

\section{Proposed Model and Estimation}\label{proposed}
In this section, we first introduce a Trivariate Bernoulli model (TBM), which is an extension of the Bivariate Bernoulli model (BBM) proposed by \citet{Chatterjee17} for modeling of the capture-recapture data in DRS. TBM will be used to incorporate the inherent dependency between capture and recapture attempts in the TRS and provide interesting interpretation of the associated model parameters.

\subsection{Trivariate Bernoulli Model (TBM)}\label{TBM}
As discussed before, three lists are prepared for estimation of the population size under TRS. It is quite natural that some individuals behave independently over the three different capture attempts and behavioral dependence may exist for the rest of the population. Let us denote $\alpha_{1}, \alpha_{2}$, and $\alpha_{3}$ be the proportion of individuals in the population for whom pair-wise dependencies between the Lists (1 and 2), (2 and 3) and (1 and 3) exist, respectively. We also consider that the second-order dependency among the List 1, List 2 and List 3 exist for $\alpha_{4}$ proportion of individuals. Therefore, the remaining (1-$\alpha$) proportion of individuals, where $\alpha=\sum_{s=1}^{4}\alpha_{s}$, behave independently over the three capture attempts. In order to capture these different dependency structures, we define a triplet ($X_{1h}^{*},X_{2h}^{*},X_{3h}^{*}$), which represents the latent capture statuses of the \textit{h}-th individual in the first, second and third attempts, respectively, for $h=1,2,\ldots,N$. The latent capture status $X_{lh}^{*}$ takes value 1 or 0, denoting the presence or absence of the \textit{h}-th individual in the $l$-th list, for $l=1,2,3$. Under this setup, for $\alpha_{1}$ proportion of individuals, the value of $X_{2h}^{*}$ is same as that of $X_{1h}^{*}$ (i.e. $X_{2h}^{*}=X_{1h}^{*}$). Similarly, $X_{3h}^{*}=X_{2h}^{*}$ for $\alpha_{2}$ proportion of individuals, $X_{3h}^{*}=X_{1h}^{*}$ for $\alpha_{3}$ proportion of individuals, and $X_{3h}^{*}=X_{2h}^{*}=X_{1h}^{*}$ for $\alpha_{4}$ proportion of individuals. Now, let us denote $Y_h$, $Z_h$, and $W_h$ as the List 1, List 2 and List 3 inclusion statuses of the \textit{h}-th individual, respectively, for $h=1,2,\ldots,N$. Note that $(Y_h,Z_h,W_h)$ is the manifestation of the latent capture statuses ($X_{1h}^{*},X_{2h}^{*},X_{3h}^{*}$) for the \textit{h}-th individual. Therefore, we can formally write the interdependence among the three lists as:
\begin{eqnarray}
(Y_h,Z_h, W_h) = \begin{cases} (X_{1h}^{*},X_{2h}^{*},X_{3h}^{*}) & \mbox{ with prob. } 1-\alpha,\\
(X_{1h}^{*},X_{1h}^{*},X_{3h}^{*})  & \mbox{ with prob. } \alpha_{1},\\
(X_{1h}^{*},X_{2h}^{*},X_{2h}^{*})  & \mbox{ with prob. } \alpha_{2},\\
(X_{1h}^{*},X_{2h}^{*},X_{1h}^{*})  & \mbox{ with prob. } \alpha_{3},\\
(X_{1h}^{*},X_{1h}^{*},X_{1h}^{*})  & \mbox{ with prob. } \alpha_{4},
\end{cases}\label{prob-model}
\end{eqnarray}  
where $X_{1h}^{*}$'s, $X_{2h}^{*}$'s and $X_{3h}^{*}$'s are independently distributed Bernoulli random variables with parameters $p_{1}$, $p_{2}$ and $p_{3}$, respectively, for all $h=1,\ldots,N$. Note that $p_{l}$ refers to the capture probability of a causally independent individual in the $l$-th list. We call this model, given in equation (\ref{prob-model}), as Trivariate Bernoulli model (TBM).  This model accounts various positive list-dependence among the capture statuses in List 1, List 2 and List 3. We define various sub-models of TBM by setting some interaction effects (i.e. $\alpha_is$) equal to zero. By construction, it is clear that the parameters of TBM possess easy interpretations with practical significance in contrast to the existing models discussed in Section \ref{existing}. 

A natural time ordering in the capture attempts is commonly found in various official statistical surveys. For time-ordered capture attempts, it is unlikely that the status in third list depends on the first list, while independent to the second list. So, one may assume $\alpha_{3}=0$, which implies pairwise dependence between first and third lists is absent. For example, in census coverage evaluation study, a third source of information, known as Administrative List Supplement (ALS), prepared prior to census operation, is often used for improving Post Enumeration Survey (PES) coverage \citep{Zaslavsky93, Wolfgang90}. We recall the model (\ref{prob-model}) with the assumption $\alpha_{3}=0$ and refer as TBM-1. Now, we denote Prob($Y=y,Z=z,W=w$) by $p_{yzw}$ for $y,z,w=\{0,1\}$. The corresponding marginal probabilities, denoted by $p_{Y}$, $p_{Z}$ and $p_{W}$, are given by
\begin{eqnarray}
p_{Y}=p_{1\cdot\cdot} & = & p_1,\nonumber\\
p_{Z}=p_{\cdot 1\cdot} & = & p_2+(p_1-p_2)(\alpha_1+\alpha_4) \text{ \vspace{20 mm} and}\nonumber\\
p_{W}=p_{\cdot\cdot 1} & = & p_3(1-\alpha)\alpha_2p_1(1-p_2).\nonumber
\end{eqnarray}	
Derivation of the aforementioned marginal probabilities associated with TBM-1 are provided in the \textit{Appendix}. Note that $p_{Y}$, $p_{Z}$ and $p_{W}$ represents the inclusion probability in List 1, List 2 and List 3, respectively.

In the context of epidemiology or public health, generally the capture attempts are not time-ordered \citep{Chao01a}. However, the capture attempts are supposed to be inter-dependent among themselves \citep{Tsay01, Ruche13} and hence, the dependence between first and third lists may be present, unlike the situation modelled by TBM-1. One can adequately model such a situation by considering $\alpha_4=0$ in (\ref{prob-model}) and we refer to the resulting model as TBM-2. Assumption of $\alpha_4=0$ implies second order interaction among three lists is absent. Therefore, marginal probabilities associated with this model are given by
\begin{eqnarray}
p_{Y}=p_{1\cdot\cdot} & = &  p_1,\nonumber\\
p_{Z}=p_{\cdot 1\cdot} & = &  \alpha_1p_1 + (1-\alpha_1)p_2\text{ \vspace{20 mm} and}\label{eqn_py=p1}\nonumber\\
p_{W}=p_{\cdot\cdot 1} & = & \alpha_3p_1 + \alpha_2p_2 + (1-\alpha_2-\alpha_3)p_3.\nonumber
\end{eqnarray}
Derivation for finding the aforementioned marginal probabilities associated with TBM-2, are presented in the \textit{Appendix}. 

As discussed, the proposed model, given by (\ref{prob-model}), accounts various positive list-dependence among the capture statuses in Lists 1, 2 and 3. Interestingly, one can easily obtain the $M_t$ model as a special case and modify the proposed model to incorporate negative dependence.
\begin{remark}\label{remark00}
	The TBM, given by (\ref{prob-model}), reduces to the $M_t$ model (i.e. there is no case of causal dependency) when $\alpha_s=0$, for all $s=1,\ldots,4$.
\end{remark} 
\begin{remark}\label{remark01}
	One can redefine the proposed TBM-1 model in order to incorporate negative dependency, useful for time-ordered capture attempts, as
	\begin{equation}
	(Y_h,Z_h, W_h) = \begin{cases} (X_{1h}^{*},X_{2h}^{*},X_{3h}^{*}) & \mbox{ with prob. } 1-\beta,\\
	(X_{1h}^{*},1-X_{1h}^{*},X_{3h}^{*})  & \mbox{ with prob. } \beta_{1},\nonumber\\
	(X_{1h}^{*},X_{2h}^{*},1-X_{2h}^{*})  & \mbox{ with prob. } \beta_{2},\nonumber
	\end{cases}\label{prob-model-aversion-time-ordered}
	\end{equation}
	where $X_{1h}^{*}$'s, $X_{2h}^{*}$'s and $X_{3h}^{*}$'s are defined as in (\ref{prob-model}) and $\beta=\beta_{1}+\beta_{2}$. On the contrary, when three attempts are not necessarily time-ordered, TBM-2 can be redefined for the same purpose as follows:
	\begin{equation}
	(Y_h,Z_h, W_h) = \begin{cases} (X_{1h}^{*},X_{2h}^{*},X_{3h}^{*}) & \mbox{ with prob. } 1-\beta,\\
	(X_{1h}^{*},1-X_{1h}^{*},X_{3h}^{*})  & \mbox{ with prob. } \beta_{1},\nonumber\\
	(X_{1h}^{*},X_{2h}^{*},1-X_{2h}^{*})  & \mbox{ with prob. } \beta_{2},\nonumber\\
	(X_{1h}^{*},X_{2h}^{*},1-X_{1h}^{*})  & \mbox{ with prob. } \beta_{3},\nonumber
	\end{cases}\label{prob-model-aversion-time-unordered}
	\end{equation}
	where $\beta=\beta_{1}+\beta_{2}+\beta_{3}$. 
\end{remark}

\subsection{Estimation}\label{estimation}
A classical approach for estimating $N$ in the context of CMR, or equivalently MRS, is based on \textit{likelihood theory}, where the data (i.e. all observed cell counts in Table \ref{Tab:1}) follow a multinomial distribution with index parameter $N$ and the associated cell probabilities $\{p_{ijk}:i,j,k=0,1;i=j=k\neq0\}$ \citep{Sanathanan72}.  
Note that the proposed TBM-1 is characterized by the parameter vector $\Theta_1=(N,\alpha_1,\alpha_2,\alpha_4,p_1,p_2,p_3)$. Therefore, using the relations between the cell probabilities $\{p_{ijk}\}$ and $(\alpha_1,\alpha_2,\alpha_4,p_1,p_2,p_3)$, as provided in the \textit{Appendix}, the likelihood function of $\Theta_1=(N,\alpha_1,\alpha_2,\alpha_4,p_1,p_2,p_3)$ under TBM-1 is given by
\begin{eqnarray}\label{L_Model_I}
L(\Theta_1) & \propto & \frac{N!}{(N-x_{0})!} [(1-\alpha)p_1 p_2 p_3 + \alpha_1 p_1 p_3 + \alpha_2 p_1 p_2 + \alpha_4 p_1]^{x_{111}}\nonumber\\
&&\times
[(1-\alpha) p_1 p_2 (1-p_3) + \alpha_1 p_1 (1-p_3)]^{x_{110}}\nonumber\\
&&\times
[(1-\alpha)(1-p_1)p_2 p_3 + \alpha_2 (1-p_1) p_2]^{x_{011}}\nonumber\\
&&\times
[(1-\alpha)p_1(1-p_2)(1-p_3) + \alpha_2 p_1 (1-p_2)]^{x_{100}}\nonumber\\
&&\times
(1-\alpha)^{x_{101}+x_{010}}\left[p_1(1-p_2)p_3\right]^{x_{101}} \left[(1-p_1)p_2(1-p_3)\right]^{x_{010}}\nonumber\\
&&\times
[(1-\alpha)(1-p_1)(1-p_2)p_3 + \alpha_1 (1-p_1) p_3]^{x_{001}}\nonumber\\
&&\times
[(1-\alpha) (1-p_1) (1-p_2) (1-p_3) + \alpha_1 (1-p_1) (1-p_3)\nonumber\\ 
&& +  \alpha_2 (1-p_1) (1-p_2) + \alpha_4 (1-p_1)]^{N-x_{0}}.\nonumber
\end{eqnarray}
However, explicit solution for MLE of $\Theta_1$ is not possible as the likelihood $L(\Theta_1)$ is mathematically intractable. One can use the Newton-Raphson method to maximize the log-likelihood for estimation of $\Theta_1$, considering $N$ as continuous parameters. Alternatively, any standard software package equipped for optimization purpose (e.g., \textit{optim} in the package R) can be used. Note that the log-likelihood function involves $log(N!)$, which may create computational difficulty for large values of $N$. In order to avoid such issues we approximate $log(N!)$ as $NlogN-N$ \citep{Wells86}. Alternatively, one can also use the R command `lgamma(N+1)'. 

Similarly, the likelihood function of $\Theta_2=(N,\alpha_1,\alpha_2,\alpha_3,p_1,p_2,p_3)$ under TBM-2 is given by
\begin{eqnarray}\label{L_Model_II}
L(\Theta_2) & \propto & \frac{N!}{(N-x_{0})!} [(1-\alpha)p_1 p_2 p_3 + \alpha_1 p_1 p_3 + \alpha_2 p_1 p_2 + \alpha_3 p_1 p_2]^{x_{111}}\nonumber\\
&&\times
[(1-\alpha) p_1 p_2 (1-p_3) + \alpha_1 p_1 (1-p_3)]^{x_{110}}\nonumber\\
&&\times
[(1-\alpha)(1-p_1)p_2 p_3 + \alpha_2 (1-p_1) p_2]^{x_{011}}\nonumber\\
&&\times
[(1-\alpha)p_1(1-p_2)(1-p_3) + \alpha_2 p_1 (1-p_2)]^{x_{100}}\nonumber\\
&&\times
[(1-\alpha)p_1(1-p_2)p_3+ \alpha_3 p_1 (1-p_2)]^{x_{101}}\nonumber\\
&&\times
\left[(1-\alpha)(1-p_1)p_2(1-p_3)+\alpha_3(1-p_1)p_2\right]^{x_{010}}\nonumber\\
&&\times
[(1-\alpha)(1-p_1)(1-p_2)p_3 + \alpha_1 (1-p_1) p_3]^{x_{001}}\nonumber\\
&&\times
[(1-\alpha) (1-p_1) (1-p_2) (1-p_3) + \alpha_1 (1-p_1) (1-p_3)\nonumber\\ 
&& +  \alpha_2 (1-p_1)(1-p_2) + \alpha_3 (1-p_1)(1-p_2)]^{N-x_{0}}.\nonumber
\end{eqnarray}
Here also, the  explicit solution for MLE of $\Theta_2$ is not possible and hence, any of the optimization methods, discussed earlier in case of MLE for $\Theta_1$, can be applied. Under some regularity conditions, consistency and asymptotic normality of the MLE of ${N}$ and the associated model parameters hold \citep{Sanathanan72}. One can also find the asymptotic variance of the estimated population size $\hat{N}$ using the observed Fisher information matrix.

\section{Real Data Examples}\label{real_data_analysis}
Application of TRS are commonly found in the domain of public health and population studies. In this section, we consider one data set from each of these aforementioned application areas. Both the datasets on Malaria incidence and census coverage study, do not show any considerable presence of heterogeneity in capture probabilities across individuals in each capture attempt \citep{Hest02,Zaslavsky93}. Hence, we apply the proposed models TBM-1 and TBM-2 (\textit{see} Section \ref{TBM}) as well as the existing competitors $M_{tb}$ and log-linear models (\textit{see} Sections \ref{ecological}-\ref{log-linear}) to the data sets for estimation of the size of respective populations based on the maximum likelihood method. As discussed in the previous Section, the \textit{optim} package in R software is used for finding the MLEs of the parameters associated with TBM-1 and TBM-2. We also observe similar results using the Newton-Raphson method for the same purpose. For estimation based on model $M_{tb}$, we present the results of unconditional MLE (i.e. UMLE), since it performs better than other two likelihood based estimation discussed in \citep{Chao00}. 

\subsection{Example I: Malaria Incidence in Netherlands}
We first analyze Malaria incidence data from Netherlands in 1996 \citep{Hest02}, which is presented in the top panel of Table \ref{Tab:6}. This data was collected based on capture-recapture technique by three incomplete and partially overlapping malaria registers: $(i)$ Notification office, $(ii)$ Hospital admission registration, and $(iii)$ Laboratory survey. Usually, in the study on infectious disease, different registration systems (e.g., notification registrar, laboratory and hospital) function in cooperation with one another, which resulted in positive dependence among the capture attempts \citep{Hest07}. Previously, \citet{Hest02} analyzed this data using LLM-2. So, we analyze this interdependence and estimate the population size applying the proposed TBM-1 and TBM-2 along with other existing models discussed in Section \ref{existing}. Findings from our analysis including relative asymptotic standard error (RASE), asymptotic confidence interval (ACI) of the estimates are presented in the top panel of Table \ref{Tab:7}. For the purpose of model selection, we also present AIC values (along with the deviance and number of free parameters) and find that TBM-1 and TBM-2 fits the data better compared to the existing competitors. It is seen that TBM-1 has slight edge over TBM-2 with respect to AIC and RASE. Under TBM-1, estimates of the dependence parameters $\alpha_{i}$s indicate $29\%$ ascertainment of the total Malaria incidences are causally dependent, whereas only $5\% $ of the total Malaria affected patients exhibit perfect positive association among the three lists.

\subsection{Example II: Census Coverage Study on Urban Adult Black Males}
Now we consider a TRS data set from a census coverage study 1990 on the population of urban adult black males in United States \citep{Zaslavsky93}, (\textit{see} the bottom panel of Table \ref{Tab:6}). This population was believed to be among the most under counted by the census and the most underestimated by the $M_t$ model based on census and post enumeration survey (PES), due to correlation bias \citep{Fay88, Chatterjee16b}. As discussed in Section \ref{sec:intro}, inclusion of an additional source comprising census and PES with the original DRS might be helpful in order to reduce the bias in the estimate. Hence, Administrative List Supplement (ALS), as a third source, plays a vital role to assess the possible positive dependence between census and PES and to obtain estimate of the population size more efficiently. Here we consider two sub-populations of urban adult black males (\textit{i}) Renters (R2) with age interval 20-29 years and (\textit{ii}) Renters (R3) with age interval 30-44 years in stratum 11. See \citet{Zaslavsky93} for detailed data description and associated analysis based on LLM-1. Results of our analysis are presented in the bottom panel of Table \ref{Tab:7}. For both the sub-populations R2 and R3, the AIC values for TBM-1 and TBM-2 are much smaller compared to the existing competitors. However, TBM-1 fits the data better compared to TBM-2 for both the sub-populations R2 and R3. Interestingly, the estimate of the population size under TBM-2 is smaller and all other estimates are larger compared to that of under TBM-1 for both the sub-populations. A similar pattern has been observed in our simulation study when the data is generated from TBM-1 with moderately large population size (\textit{see} top panel of Table \ref{Tab:3}). From the estimates of dependence parameters $\alpha_i$'s, one can infer that 56\% and 79\% individuals in the sub-population R2 and R3, respectively, are causally dependent. We also find that 21\% and 23\% individuals, in R2 and R3, respectively, possess perfect positive association among the three lists.

\begin{table}[ht]
	\small
	\centering
	\caption{Data sets used for illustration of the proposed and existing models.}
	\resizebox{5.6in}{!}{
		\begin{tabular}{|lcccccccc|}
			\hline
			Data set & $x_{111}$ & $x_{110}$ & $x_{101}$ & $x_{011}$ & $x_{100}$ & $x_{010}$ & $x_{001}$ & Total ($x_0$)\\
			\hline
			\hline
			\multicolumn{9}{|c|}{}\\
			Malaria Incidents & 123 & 127 & 94 & 37 & 189 & 41 & 54 & 665 \\
			\multicolumn{9}{|c|}{}\\
			Urban Black Males & \multicolumn{8}{c|}{}\\
			\hspace{3mm} Renters R2 (20-29 years) & 58 & 69 & 12 & 11 & 41 & 34 & 43 & 268 \\
			\hspace{3mm} Renters R3 (30-44 years) & 72 & 69 & 7 & 13 & 32 & 13 & 43 & 249\\
			\hline
		\end{tabular}
	}
	\label{Tab:6}
\end{table}

\begin{table}[h]
	\small
	\centering
	\caption{Summary results of the data analyses of proposed TBM-1 and TBM-2 models along with the existing models.}
	\begin{minipage}{20cm}
		\begin{tabular}{|lrccccc|}
			\hline
			& &  & & &  & \\
			Data set & & TBM-1 & TBM-2 & LLM-1 & LLM-2 & $M_{tb}$\\
			\hline
			\hline
			& & & & & & \\
			Malaria Incidents & $\hat{N}$(RASE) & 775 (0.034) & 798 (0.113) & 781 (0.046) & 774 (0.029) & 813 (0.115)\\
			&  ACI & (723, 827) & (621, 975) & (711, 851) & (729, 817)  & (630, 996)\\
			& No. of free parameters & 7 & 7 & 7  & 6 & 5\\
			& Deviance\footnote{Deviance=-$2 \times \log(\text{Likelihood})$} & 0.008 & 0.049 &  43.637 & 43.711 & 15.567\\
			& AIC\footnote{AIC=Deviance + $2 \times(\text{No. of free parameters})$} & 14.008 & 14.049 &  57.637 & 55.711 & 25.567\\
			& & & & & & \\
			\multicolumn{7}{|l|}{Urban Black Males}\\
			R2: Renters &  $\hat{N}$(RASE) & 474 (0.153)& 364 (0.082) & 649 (0.121) & 414 (0.131) & 597 (0.321)\\
			&  ACI & (332, 616) & (305, 423) & (495, 803) & (308, 520)  & (221, 973)\\
			& No. of free parameters & 7 & 7 & 7  & 6 & 5 \\
			& Deviance & 0.002 & 10.769 & 37.091  & 43.609 & 22.909\\
			&  AIC & 14.002 & 24.769 & 51.091 & 55.609  & 32.909\\
			& & & & & & \\
			R3: Renters & $\hat{N}$(RASE) & 449 (0.223) & 319 (0.077) & 454 (0.268) & 445 (0.198) & 550 (0.256)\\
			&  ACI & (253, 645) & (271, 367) & (216, 692) & (272, 618)  & (274, 826)\\
			& No. of free parameters & 7 & 7 & 7  & 6 & 5\\
			& Deviance & 0.002 & 4.349 & 35.742  & 35.752 & 36.385\\
			& AIC & 14.002 & 18.349 & 49.742 & 47.752  & 46.385\\
			\hline
		\end{tabular}
	\end{minipage}
	\label{Tab:7}
\end{table}

\section{Simulation Study}\label{Simulation}
In this section, we perform extensive simulation study for comparing the performances of the proposed estimators based on TBM models (Section \ref{proposed}) with the existing competitors discussed in Section \ref{existing}. First, we consider 5 different choices of parameter values P1-P5 (\textit{see} upper panel of Table \ref{Tab:2}) and P6-P10 (\textit{see} lower panel of Table \ref{Tab:2}), under each of the two simulation models TBM-1 and TBM-2, respectively, for $N=400, 1000$. The expected numbers of distinct captured individuals given the population size, denoted by $E\left[ x_0|N \right]$, are also presented in the last two columns of Table \ref{Tab:2} for each of the simulation models. Note that $M_{tb}$ and log-linear models are not suitable for generating a TRS data structure (incomplete $2^3$ table) as presented in Table \ref{Tab:1}. To compute the average estimate ($\hat{N}$), relative root mean square error (RRMSE), and $95\%$ confidence interval (CI) of $N$, we first draw a sample from the proposed model (TBM-1 or TBM-2) using the genesis provided in (\ref{prob-model}), and present the data in the form of Table \ref{Tab:1}. Each competitive model is then fitted to the simulated data to obtain an estimate of the population size $N$. This process is repeated $5000$ times. Based on these $5000$ estimates, we compute the average estimate ($\hat{N}$), and RRMSE of $N$ for each of the competitive models. We also find the $2.5$th and the $97.5$th percentiles by ordering these replicated estimates to construct a $ 95 \%$ confidence interval of $N$. Here also, we used \textit{optim} package in R software for finding MLEs of the model parameters, and the results are similar to the Newton-Raphson method. The findings are summarized below. 

As expected, the estimator derived from the true model performs best in terms of bias and RRMSE. Interestingly, TBM-2 outperforms all the existing competitors (LLM-1, LLM-2 and $M_{tb}$), when the data is generated from TBM-1. However, TBM-2 underestimates and the log-linear models overestimate the population size for all choices of parameters values under TBM-1. On the contrary, no such pattern is observed when the data is generated from TBM-2. Further from Tables \ref{Tab:3} and \ref{Tab:4}, it is also observed that RRMSE values decrease when true population size $N$ increases from 400 to 1000 in case of all simulated populations under both TBM-1 and TBM-2. It is also observed from the analysis that our proposed models produce 95\% confidence intervals for $N$ with shortest length than that of other competitive models in the study. Specifically for populations simulated from TBM-1, log-linear models produce confidence intervals with very wide length (most of the estimated confidence interval length are more than $500$).

\begin{table}[h]
	\small
	\caption{Composition of simulated populations.}
	\label{Tab:2}
	\begin{center}
		\begin{tabular}{|cccccc|}
			\hline
			\hline
			Simulation Model & Population & ($p_{1}$, $p_{2}$, $p_{3}$) & ($\alpha_{1}$, $\alpha_{2}$, $\alpha_{3}$, $\alpha_{4}$) & $E[x_0|$N=400$]$ & $E[x_0|$N=1000$]$\\
			\hline
			TBM-1 &	P1  & ($0.4$, $0.5$, $0.6$) & ($0.6$, $0.1$, 0, $0.2$) & 278  &  $694$\\
			&   P2  & ($0.4$, $0.5$, $0.6$) & ($0.2$, $0.6$, 0, $0.1$) & 280 &  $700$\\
			&	P3  & ($0.6$, $0.7$, $0.6$) & ($0.4$, $0.1$, 0, $0.4$) & 304  & $776$ \\
			&   P4  & ($0.6$, $0.4$, $0.5$) & ($0.5$, $0.3$, 0, $0.1$) &  310 &  $776$\\
			&   P5  & ($0.6$, $0.4$, $0.5$) & ($0.3$, $0.3$, 0, $0.3$) & 294  &  $736$\\
			\hline
			TBM-2 &   P6 & ($0.4$, $0.5$, $0.6$) & ($0.6$, $0.1$, 0.2, $0$) & 302  & 754  \\
			&   P7 & ($0.4$, $0.5$, $0.6$) & ($0.2$, $0.6$, 0.1, $0$) &  $292$ &  $730$\\
			&   P8 & ($0.6$, $0.7$, $0.6$) & ($0.4$, $0.1$, 0.4, $0$) & 349  & 871 \\
			&   P9 & ($0.6$, $0.4$, $0.5$) & ($0.5$, $0.3$, 0.1, $0$) & 317  & 792 \\
			&   P10 & ($0.6$, $0.4$, $0.5$) & ($0.3$, $0.3$, 0.3, $0$) & 314  & 784 \\
			\hline
			\hline
		\end{tabular}
	\end{center}
\end{table}

\begin{table}[h]
	\small
	\caption{Summary results on the estimators of $N$ under the simulated model \textbf{TBM-1} with $N=400, 1000$.}
	\label{Tab:3}
	\begin{center}
		\begin{minipage}{14cm}
			\begin{tabular}{|lrccccc|}
				\hline
				Model & & P1 & P2 & P3 & P4 & P5\\
				\hline
				\multicolumn{7}{|c|}{$N=400$}\\
				$\textbf{TBM-1}$  & $\hat{N}$(RRMSE) & 405 (0.083) & 408 (0.095) & 409 (0.116) & 414 (0.120) & 408 (0.102)\\
				& CI  & $(394, 512)$& (392, 537) & $(332, 554)$ & $(386, 569)$ & $(366, 546)$\\
				$\textbf{TBM-2}$  & $\hat{N}$(RRMSE) & 344 (0.146) & 363 (0.098) & 304 (0.242) &  375 (0.067) & 294 (0.265)\\
				& CI  & $(315, 372)$& (336, 393) & $(286, 320)$ & $(356, 396)$ & $(277, 531)$\\
				$\textbf{LLM-1}$  & $\hat{N}$(RRMSE) & 461 (0.482) & 461 (0.497) & 442 (0.417) & 415 (0.261) & 423 (0.324)\\
				&  CI  & $(294, 975)$ &$(296, 978)$ & $(312, 884)$  & $(312, 680)$ & $(297, 771)$ \\
				$\textbf{LLM-2}$  & $\hat{N}$(RRMSE) & 557 (0.661) & 783 (1.336) & 432 (0.278) & 687 (0.934) & 535 (0.522)\\
				&  CI  & $(350, 1185)$ & $(437, 2009)$ & $(329, 766)$ & $(458, 1270)$ & $(383, 949)$\\
				\textbf{$M_{tb}$}  & $\hat{N}$(RRMSE) & 551 (0.468)& 332 (0.207) & 499 (0.250)  & 448 (0.193) & 505 (0.264)\\
				&  CI  & $(400, 736)$& $(278, 400)$ & $(481, 518)$  & $(399, 574)$ & $(503, 508)$\\
				\multicolumn{7}{|c|}{}\\
				\multicolumn{7}{|c|}{$N=1000$}\\
				$\textbf{TBM-1}$  & $\hat{N}$(RRMSE) & 1001 (0.020) & 1001 (0.014) & 1000 (0.007) & 1002 (0.019) & 999 (0.025)\\
				& CI  & $(995, 1005)$ & $(997, 1003)$ & $(999, 1001)$  & $(995, 1006)$ & $(986, 1009)$\\
				$\textbf{TBM-2}$  & $\hat{N}$(RRMSE) & 694 (0.306) & 974 (0.067) & 759 (0.241) & 985 (0.023) & 736 (0.264)\\
				& CI  & $(666, 722)$& (701, 1019) & $(733, 786)$ & $(953, 1021)$ & $(708, 763)$\\
				$\textbf{LLM-1}$  & $\hat{N}$(RRMSE) & 1076 (0.236) & 1066 (0.215) & 1037 (0.179) & 999 (0.116) & 1011 (0.144)\\
				&  CI  & $(813, 1678)$ & $(819, 1569)$ & $(833, 1467)$ & $(837, 1285)$ & $(815, 1359)$\\
				$\textbf{LLM-2}$  & $\hat{N}$(RRMSE) & 1298 (0.397) & 1765 (0.869) & 1029 (0.133) & 1611 (0.664) & 1271 (0.319)\\
				&  CI  & $(974, 1977)$ & $(1257, 2796)$ & $(873, 1343)$ & $(1254, 2225)$ & $(1036, 1687)$\\
				\textbf{$M_{tb}$}  & $\hat{N}$(RRMSE) & 1301 (0.338) & 901 (0.156) & 1106 (0.108) & 1033 (0.074) & 865 (0.149)\\
				&  CI  & $(999, 1578)$ & $(715, 1000)$ & $(1078, 1137)$ & $(999, 1212)$ & $(810, 1029)$\\
				\hline
			\end{tabular}
		\end{minipage}
	\end{center}
\end{table}

\begin{table}[h]
	\small
	\caption{Summary results on the estimators of $N$ under the simulated model \textbf{TBM-2} with $N=400, 1000$.}
	\label{Tab:4}
	\begin{center}
		\begin{minipage}{14cm}
			\begin{tabular}{|lrccccc|}
				\hline
				Model & & P6 & P7 & P8 & P9 & P10\\
				\hline
				\multicolumn{7}{|c|}{$N=400$}\\
				$\textbf{TBM-1}$  & $\hat{N}$(RRMSE) &  442 (0.167) & 425 (0.163) & 349 (0.129)  & 407 (0.074) & 376 (0.079)\\
				& CI  & $(384, 558)$& (396, 605) & $(336, 361$  & $(365, 497)$ & $(340, 416)$\\
				$\textbf{TBM-2}$ & $\hat{N}$(RRMSE) & 396 (0.052) & 395 (0.052) & 399 (0.032) & 399 (0.025) & 395 (0.039)\\
				& CI  & $(356, 442)$& (351, 437) & $(375, 426)$ & $(376, 423)$ & $(350, 418)$\\
				$\textbf{LLM-1}$  & $\hat{N}$(RRMSE) & 480 (0.327)& 470 (0.330) & 479 (0.269) & 397 (0.115) & 390 (0.098)\\
				&  CI  & $(352, 749)$ & (343, 750) & $(384, 658)$ & $(335, 514)$ & $(337, 481)$\\
				$\textbf{LLM-2}$  & $\hat{N}$(RRMSE) & 354 (0.129) & 448 (0.204) & 361 (0.099) & 427 (0.112) & 343 (0.147)\\
				&  CI  & $(316, 408)$ & (363, 612) & $(345, 379)$ & $(371, 514)$ & $(320, 367)$\\
				\textbf{$M_{tb}$}  & $\hat{N}$(RRMSE) & 648 (0.637)& 308 (0.235) & 470 (0.180)  & 447 (0.140) & 429 (0.087)\\
				&  CI  & $(561, 773)$& $(283, 350)$ & $(434, 518)$  & $(421, 550)$ & $(358, 449)$\\
				\multicolumn{7}{|c|}{}\\
				\multicolumn{7}{|c|}{$N=1000$}\\
				$\textbf{TBM-1}$  & $\hat{N}$(RRMSE) & 754 (0.246) & 1034 (0.086) & 871 (0.128)  & 998 (0.033) & 971 (0.051)\\
				& CI  & $(728, 781)$ & (996, 1285) & $(850, 891)$ & $(923, 1093)$ & $(881, 1025)$\\
				$\textbf{TBM-2}$  & $\hat{N}$(RRMSE) & 997 (0.025) & 997 (0.017) & 998 (0.021) & 999 (0.009) &  995 (0.023)\\
				& CI  & $(922, 1047)$ & $(951, 1014)$ & $(953, 1040)$ & $(973, 1011)$ & $(908, 1016)$\\
				$\textbf{LLM-1}$  & $\hat{N}$(RRMSE) & 1165 (0.217) & 1135 (0.195) & 1172 (0.200) & 977 (0.068) & 964 (0.065)\\
				&  CI  & $(950, 1491)$ & $(926, 1474)$ & $(1018, 1409)$ & $(876, 1125)$ & $(879, 1088)$\\
				$\textbf{LLM-2}$  & $\hat{N}$(RRMSE) & 882 (0.123) & 1101 (0.134) & 902 (0.099) & 1058 (0.079) & 856 (0.145)\\
				&  CI  & $(820, 957)$ & $(961, 1303)$ & $(876, 930)$ & $(967, 1175)$ & $(822, 893)$\\
				\textbf{$M_{tb}$}  & $\hat{N}$(RRMSE) & 1120 (0.222) & 864 (0.183) & 1149 (0.150) & 1082 (0.094) & 1053 (0.062)\\
				&  CI  & $(1004, 1605)$ & $(728, 1028)$ & $(1125, 1206)$ & $(1043, 1227)$ & $(935, 1084)$\\
				\hline
			\end{tabular}
		\end{minipage}
	\end{center}
	\label{sim_2}
\end{table}

\section{Concluding Remarks}\label{conclusion}
In this article, we address the issue of population size estimation incorporating the possible dependence among the capture and recapture attempts under TRS. For this purpose, we introduce a Trivariate Bernoulli model (TBM), possibly for the first time, and obtain estimate of the population size as well as dependence parameters under two different realistic scenarios. Though our proposed models account for positive dependence, one can easily modify and apply the TBM for modeling negative dependence (\textit{see} Remark \ref{remark01}). The proposed models seem to have an edge in terms of ease of interpretation and have much wider domain of applicability in the fields of public health, demography and social sciences. From the analysis of real data, one can observe that the TBMs exhibit remarkable improvement over the existing models. Further, this model successfully incorporates second order interaction among the three lists in TRS, unlike the existing $M_{tb}$ and log-linear models. Not only the estimate of the population size $N$, estimates of the associated dependence parameters $\hat{\alpha}_is$, provide specific insights into the capture-recapture mechanism. As discussed in Section \ref{sec:intro}, `individual heterogeneity' i.e. heterogeneity in the capture probabilities may exist across individuals in each list, especially when suitable post-stratification is not possible. \citet{WorkingGroup95} presented a nice overview of the log-linear models for accounting dependency among the lists with heterogeneity in the capture probabilities. It will be an interesting problem to model such data using TBM and develop associated estimation methodology. It would also be interesting to develop test procedures for significance of two-factor and three-factor interactions among the different capture attempts in TRS and this may be a worthwhile topic for future research. Further, one can easily extend the proposed models for MRS.

\clearpage
\section*{Appendix: Derivation of Marginal Capture Probabilities in three lists}\label{appendix}
\subsection*{Derivation of Marginal Capture Probabilities under the model TBM-1}
The cell probabilities, except $p_{000}$, corresponding to Table \ref{Tab:1} under TBM-1 are given by
\begin{eqnarray}
p_{111} & = & (1-\alpha)p_1p_2p_3+\alpha_1p_1p_3+\alpha_2p_1p_2+\alpha_4p_1,\nonumber\\
p_{101} & = & (1-\alpha)p_1(1-p_2)p_3,\nonumber\\
p_{110} & = & (1-\alpha)p_1p_2(1-p_3)+\alpha_1p_1(1-p_3),\nonumber\\
p_{100} & = & (1-\alpha)p_1(1-p_2)(1-p_3)+\alpha_2p_1(1-p_2),\nonumber\\
p_{011} & = & (1-\alpha)(1-p_1)p_2p_3+\alpha_2(1-p_1)p_2,\nonumber\\
p_{001} & = & (1-\alpha)(1-p_1)(1-p_2)p_3+\alpha_1(1-p_1)p_3,\nonumber\\
p_{010} & = & (1-\alpha)(1-p_1)p_2(1-p_3).\nonumber
\end{eqnarray}
Therefore, the three marginal probabilities associated with the inclusion status of an individual in the three lists are given by
\begin{eqnarray}
p_{Y}=p_{1\cdot\cdot}=(p_{111}+p_{110}+p_{101}+p_{100}) & = & p_1,\nonumber\\
p_{Z}=p_{\cdot 1\cdot}=(p_{111}+p_{110}+p_{011}+p_{010}) & = & p_2+(p_1-p_2)(\alpha_1+\alpha_4) \text{ \vspace{20 mm} and}\nonumber\\
p_{W}=p_{\cdot\cdot 1}=(p_{111}+p_{101}+p_{011}+p_{001}) & = & p_3(1-\alpha)\alpha_2p_1(1-p_2).\nonumber
\end{eqnarray}

\subsection*{Derivation of Marginal Capture Probabilities under the model TBM-2}	
Similarly, in the context of TBM-2, all the cell probabilities corresponding to Table \ref{Tab:1}, except $p_{000}$, are given as
\begin{eqnarray}
p_{111} & = & (1-\alpha)p_1p_2p_3+\alpha_1p_1p_3+(\alpha_2+\alpha_3)p_1p_2,\nonumber\\
p_{101} & = & (1-\alpha)p_1(1-p_2)p_3+\alpha_3p_1(1-p_2),\nonumber\\
p_{110} & = & (1-\alpha)p_1p_2(1-p_3)+\alpha_1p_1(1-p_3),\nonumber\\
p_{100} & = & (1-\alpha)p_1(1-p_2)(1-p_3)+\alpha_2p_1(1-p_2),\nonumber\\
p_{011} & = & (1-\alpha)(1-p_1)p_2p_3+\alpha_2(1-p_1)p_2,\nonumber\\
p_{001} & = & (1-\alpha)(1-p_1)(1-p_2)p_3+\alpha_1(1-p_1)p_3,\nonumber\\
p_{010} & = & (1-\alpha)(1-p_1)p_2(1-p_3)+\alpha_3(1-p_1)p_2.\nonumber
\end{eqnarray}
Therefore, the three marginal probabilities corresponding to TBM-2 are given by
\begin{eqnarray}
p_{Y}=p_{1\cdot\cdot}=(p_{111}+p_{110}+p_{101}+p_{100}) & = &  p_1,\nonumber\\
p_{Z}=p_{\cdot 1\cdot}=(p_{111}+p_{110}+p_{011}+p_{010}) & = &  \alpha_1p_1 + (1-\alpha_1)p_2\text{ \vspace{20 mm} and}\label{eqn_py=p1}\nonumber\\
p_{W}=p_{\cdot\cdot 1}=(p_{111}+p_{101}+p_{011}+p_{001}) & = & \alpha_3p_1 + \alpha_2p_2 + (1-\alpha_2-\alpha_3)p_3.\nonumber
\end{eqnarray}

\bibliographystyle{apalike}
\bibliography{Bibliography_TBM}

\end{document}